\documentclass{PoS}

\title{Complex 2D Matrix Model and Its Application to $N_c$-dependence of Hadron Structures}

\ShortTitle{Complex 2D Matrix Model and Its Application to $N_c$-dependence of Hadron Structures}

\author{\speaker{Kanabu Nawa}\\
        Theoretical Research Division, Nishina Center, RIKEN, Saitama 351-0198, Japan\\
        E-mail: \email{knawa@riken.jp}}

\author{Sho Ozaki\\
        Institute of Physics and Applied Physics, Yonsei University,
        Seoul 120-749, Korea\\
                                            }

\author{Hideko Nagahiro\\
        Department of Physics, Nara Women's University, Nara 630-8506,
        Japan\\
                  }

\author{Daisuke Jido\\
        Department of Physics, Tokyo Metropolitan University, Hachioji,
        Tokyo 192-0397, Japan\\
                                             }

\author{Atsushi Hosaka\\
        Research Center for Nuclear Physics (RCNP), Osaka University,
        Osaka 567-0047, Japan\\
                                             }

\abstract{
We study 
the internal structure of resonance
states in a complex 2D matrix model.
We show that the geometry with ``exceptional points'' in
the complex-parameter space can be useful to discuss 
parameter dependence of the structures 
within real-parameter subspace.
By applying the model to hadron physics,
we consider the $N_c$-dependence of hadron structures
from the geometry on the complex-$N_c$ plane.
}

\FullConference{XV International Conference on Hadron Spectroscopy-Hadron 2013\\
		4-8 November 2013\\
		Nara, Japan }

\begin{document}

\section{Introduction}

\begin{figure}[t]
  \begin{center}
       \resizebox{135mm}{!}{\includegraphics{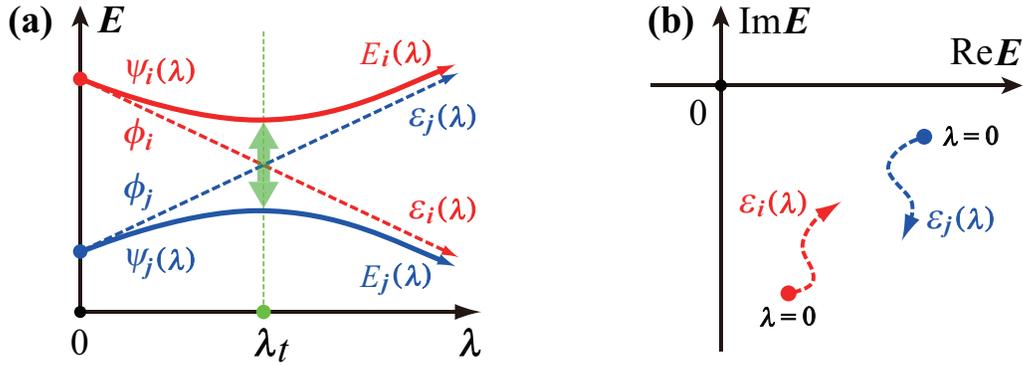}}\\
  \end{center}
\caption{(Color) (a) 
Hermitian case with
anticrossing between $i$th and $j$th eigenstates 
with variation of $\lambda\in\mbox{${\bf R}$}$.
Indices of lines are explained in the text.
(b) Non-Hermitian case of $\varepsilon_i(\lambda)$ and $\varepsilon_j(\lambda)$
with variation of 
$\lambda\in\mbox{${\bf R}$}$ on complex energy plane.
}
  \label{fig_1D_2D}
\end{figure}

The parameter dependence of quantum states 
has been extensively studied in 
various physics.
An interesting example is the solar neutrino conversion:
electron neutrinos produced in the center of the sun
via nuclear fusion reaction can become muon neutrinos
around the surface of the sun~\cite{Be}, so that less amount of electron
neutrinos
can be observed by a water Cherenkov detector on the earth, formerly known as 
``solar neutrino puzzle''~\cite{RDavis}.
To analyze such parameter dependence,
a Hermite model Hamiltonian
$\hat{H}(\lambda)$ with a real parameter $\lambda$
is usually adopted.
Let us assume that 
the eigenstates 
$\phi_i$ ($i=1,2,\cdots$)
of $\hat{H}(\lambda)$ at $\lambda=0$
can have clear characters.
The full eigenstates $\psi_i(\lambda)$ ($i=1,2,\cdots$)
of $\hat{H}(\lambda)$ 
coincide with $\phi_i$ at $\lambda=0$, i.e., $\psi_i(0)=\phi_i$.
The Neumann-Wigner non-crossing rule~\cite{NW} tells us that,
if the energy expectation values 
$\varepsilon_i(\lambda)\equiv \langle \phi_i |\hat{H}(\lambda)|\phi_i \rangle$
($i=1,2,\cdots$) cross with each other at 
$\lambda=\lambda_t\in \mbox{${\bf R}$}$,
the energy eigenvalues $E_i(\lambda)$ of 
$\psi_i(\lambda)$ can
have anticrossing at $\lambda_t$ (see Fig.~\ref{fig_1D_2D}(a)).
At this point,
the overlap of $\psi_i(\lambda)$ with $\phi_i$ is found to be exceeded by that
with $\phi_j$ ($i\neq  j$)
as $|\langle\phi_i|\psi_i\rangle|^2\leq |\langle\phi_j|\psi_i\rangle|^2$.
That is, due to orthogonality,  $\psi_i(\lambda)$ and $\psi_j(\lambda)$ 
exchange their characters
in terms of $\phi_i$ and $\phi_j$ at $\lambda=\lambda_t$.
We call such phenomena as ``nature transition''. 
In fact, 
knowing {\it a priori} the critical value $\lambda_t$
is very useful to inclusively discuss 
the parameter dependence of  
the internal structure of the quantum states.

In our work, we consider the 
open quantum systems having dissipation into 
decay channels outside of the model space.
These systems can be effectively described by
non-Hermite model Hamiltonian $\hat{H}(\lambda)$
accompanying 
with ``complex energy eigenvalues'' for ``resonance states''.
In this case,
$\varepsilon_i(\lambda)$ can move 
two dimensionally on the complex energy plane (see Fig.~\ref{fig_1D_2D}(b))
without having degeneracy at a certain value of $\lambda$.
Therefore the criterion of nature transition between resonance states
becomes completely uncertain. 
To solve this problem,
we formulate the 
complex two-dimensional (2D) matrix model. 
(Two dimensions represent two levels of resonances.)
We found that the geometry with ``exceptional points''
in the complex-parameter space can play a key role
for such parameter dependence within real-parameter subspace~\cite{KNawa}.

We apply the model to the hadron physics with $1/N_c$ expansion.
We discuss the typical $N_c$-dependence of the internal structures 
of hadrons from the geometry on the complex-$N_c$ plane~\cite{KNawa}.

\section{Complex 2D matrix model}
First we consider a two-level problem
in an open quantum system.
We describe resonance states 
in {\it bi-orthogonal representation}
as $|\phi_i)$$(i=1,2)$, 
where its bra-state is the complex conjugate 
of the Dirac bra-state $(\phi_i|\equiv\langle \phi_i^*|$~\cite{Ho}.
%
We assume that $|\phi_i)$, the eigenstates of $\hat{H}(\lambda)$ 
at $\lambda=0$,
are an appropriate basis with clear characters.
Then we consider the Hamilton matrix 
in this basis:
\begin{eqnarray}
H(\lambda)=\left(\begin{array}{cc}
                    (\phi_1|\hat{H}(\lambda)|\phi_1) &  (\phi_1|\hat{H}(\lambda)|\phi_2)      \\
                    (\phi_2|\hat{H}(\lambda)|\phi_1) &  (\phi_2|\hat{H}(\lambda)|\phi_2) 
                             \end{array}\right)\equiv
                  \left(\begin{array}{cc}
                   \varepsilon_1(\lambda) & V_{12}(\lambda) \\
                    V_{21}(\lambda)               & \varepsilon_2 (\lambda)
                             \end{array}\right).
\label{H_1}
\end{eqnarray}
$\varepsilon_i(\in\mbox{${\bf C}$})$ is the energy of $|\phi_i)$ 
and $V_{ij}(\in \mbox{${\bf C}$})$ are the interaction
satisfying
$V_{ij}(0)=0$.
$\lambda (\in\mbox{${\bf R}$})$ is a parameter,
controlling the development of the two eigenstates 
$|\psi_i(\lambda))$
which can be decomposed by $|\phi_i)$ as
$|\psi_i(\lambda)) \equiv C_{i 1}(\lambda)|\phi_1)+C_{i 2}(\lambda)|\phi_2)$ ($i=1,2$).
$C_{ij}(\lambda)$ 
carry the information about
the internal structure of the eigenstates $|\psi_i(\lambda))$ 
in terms of $|\phi_i)$.
%
%
Due to bi-orthogonality,
the norms $(\psi_i|\psi_i)=C_{i1}^2+C_{i2}^2$ can become complex, while
we simply assume the module, $|C_{ij}(\lambda)|^2$,  
to be the {\it component weights}, being suitable for narrow resonances. 

\begin{figure}[t]
  \begin{center}
\vspace*{-2mm}
       \resizebox{59mm}{!}{\includegraphics{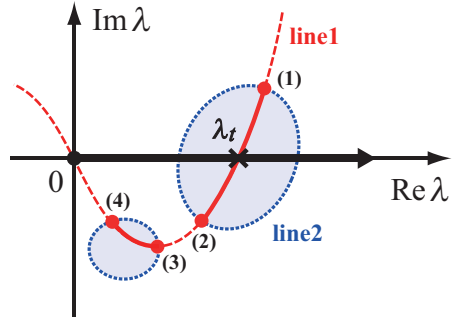}}\\
  \end{center}
\vspace{-5mm}
\caption{(Color) (a) Schematic figure of geometrical map with
transition lines and 
exceptional points 
on complex $\lambda$ plane.
Roles of Line 1 and shaded area with boundary of line 2
are given in the text.
Points (n) denote the exceptional points $\lambda_{\rm EX}^{(n)}$.
Transition lines are shown by the solid curves, satisfying the
 transition condition: $|C_{i 1}(\lambda)|^2=|C_{i 2}(\lambda)|^2$.
}
  \label{fig_1}
\end{figure}

Now let us consider the condition of the nature transition, i.e., 
$|C_{i 1}(\lambda)|^2=|C_{i 2}(\lambda)|^2$,
where the two basis components $|\phi_1)$ and $|\phi_2)$
are equally mixed 
in each eigenstate $|\psi_i(\lambda))$
as a character exchanging point.
We found that the {\it geometry on complex-$\lambda$ plane}
gives simple criterion of nature transition within the real-$\lambda$
subspace as follows.
The transition condition $|C_{i 1}(\lambda)|^2=|C_{i 2}(\lambda)|^2$
is equal to the two conditions:
\begin{eqnarray}
&&{\rm Re}[A(\lambda)^*\overline{V}(\lambda)]=0,\label{condF1} \\
&&|A(\lambda)|^2\leq|\overline{V}(\lambda)|^2, \label{condF2}
\end{eqnarray}
where $A(\lambda)\equiv
\{\varepsilon_1(\lambda)-\varepsilon_2(\lambda)\}/2$ and
$\overline{V}(\lambda)^2\equiv V_{12}(\lambda)V_{21}(\lambda)$.
The ``transition line'' is defined 
as the region satisfying 
$|C_{i 1}(\lambda)|^2=|C_{i 2}(\lambda)|^2$, i.e.,
both conditions (\ref{condF1}) and (\ref{condF2})
on the complex-$\lambda$ plane.
The line (\ref{condF1}) is named as ``line 1'', 
which 
includes the transition line.
The boundary of the region (\ref{condF2})
is named as ``line 2'', which
selects the proper part for the transition line from line~1.
%
%
If the transition line crosses the real-$\lambda$ axis,
the nature transition occurs at the crossing point  
$\lambda=\lambda_t\in \mbox{\bf R}$
(see Fig.~\ref{fig_1}).

We can show that {\it all} of the crossing points between
line 1 and 2 correspond to the ``exceptional points''~\cite{Kato},
where two energy eigenvalues coincides with each other.
The significance of the exceptional points has been
investigated
in the context of quantum chaos~\cite{QC},
where the dense exceptional points on the complex-$\lambda$ plane
indicates the development of quantum chaos in the energy-level statistics.
In fact, the exceptional point corresponds to the phase singular point of
the Berry phase~\cite{Berry}.
In our study, we can newly show that
line 1 and 2 cross with each other at all exceptional points,
so that these points
should always be the {\it end points}
of the transition lines.
In this way, the exceptional points tell us not only
the global information like development of quantum chaos,
but also the information about internal structure of each quantum level.


\section{$N_c$-dependence of hadron structures}

We employ the complex 2D matrix model (\ref{H_1}) to 
the analysis of the $N_c$-dependence of the internal structure of hadrons.
For a demonstration, we consider
the admixed nature of 
the $a_1(1260)$ meson carrying
$q\bar{q}$ and $\pi\rho$-molecule components.

First, we analyze
the scattering equation for the $\pi$-$\rho$ propagator in the 
$J^P=1^+$ channel.
Then, by reducing the relativistic eigenvalue equation to the Schr\"{o}dinger
equation of the model (\ref{H_1}) in a non-relativistic approximation,
we can get the geometrical map
on the complex-$N_c$ plane for the $a_1$ meson~\cite{KNawa}.
The  complex 2D matrix model for $a_1$ meson with
$N_c$ dependence factored out by $\lambda$ becomes
\begin{eqnarray}
H(\lambda)=
                  \left(\begin{array}{cc}
                   \frac{1}{\lambda^2}\sqrt{s_p} &  \frac{\lambda}{2\tilde{m}}\sqrt{Z}v_{a_1\pi\rho} \\
                     \frac{\lambda}{2\tilde{m}} \sqrt{Z}v_{a_1\pi\rho}  & m_{a_1}
                             \end{array}\right), \label{H_C2Da1}\\
\lambda\equiv \sqrt[4]{3/N_c}.\hspace{15mm}
\end{eqnarray}
$s_p$ is the pole mass of $\pi\rho$-molecule state and
$Z$ is its pole residue.
$v_{a_1\pi\rho}$ is a three-point coupling and 
$\tilde{m}\equiv \sqrt{\sqrt{s_p}m_{a_1}}$ is the typical mass scale of
the problem. 
These constants are numerically estimated by perturbative resummation
for the chiral Lagrangian
induced by holographic QCD~\cite{SS,NSK} as 
$\sqrt{s_p}\simeq 1012-221i$,
$\sqrt{Z}\simeq 84-21i$,
$v_{a_1\pi\rho}\simeq -6493$, and 
$m_{a_1}=1189$ in MeV unit. 

\begin{figure}[t]
  \begin{center}
  \hspace*{-2mm}
       \resizebox{120mm}{!}{\includegraphics{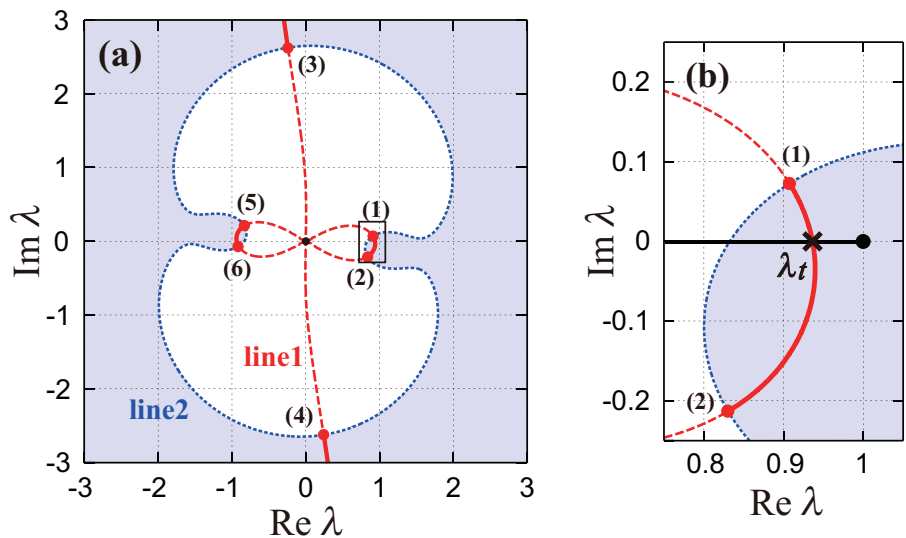}}\\
  \end{center}
\vspace{-5mm}
\caption{(Color) (a) Geometrical map
on the complex-$N_c$ plane with $\lambda=\sqrt[4]{3/N_c}$.
(b) Close-up figure around 
a blank square in (a). }
\label{fig_2}
\end{figure}

By applying the conditions (\ref{condF1}) and (\ref{condF2})
to the Hamiltonian (\ref{H_C2Da1}),
we can figure out the geometrical map on the complex-$N_c$ plane for the $a_1$ meson in Fig.~\ref{fig_2}.
%
%
%
The transition line as the solid curve 
is found to cross the real $\lambda$ axis between
$\lambda=0$ ($N_c=\infty$) and $\lambda=1$ ($N_c=3$).
%
A critical color number of the nature transition
can be calculated by the crossing point
as $\lambda_t=\sqrt[4]{3/N_c}\sim 0.93$, i.e., $N_c\sim 4.0$.
%
%
That is, with continuous variation of $N_c$ from
$\infty$ to $3$,
the internal structures of two hadronic states are exchanged
in terms of appropriate basis $q\bar{q}$ and $\pi\rho$-molecule at the 
critical color number: $N_c\sim 4.0$.
Such a critical color number with character exchange for the $a_1$ meson
is also suggested by analyzing
the pole residues 
in Ref.~\cite{NNOJH}.
In this way,
by taking into account the existence of nature transition 
from the geometry on the complex-$N_c$ plane,
we can 
discuss the typical $N_c$-dependence of the hadron structures
from $N_c=\infty$ to $3$.

\section{Summary}

In this work, we discuss the 
parameter-dependence of the internal structure of resonances
from the geometry of a complex-parameter space.
By applying the model to hadron physics with $1/N_c$ expansion,
we consider the typical $N_c$-dependence of hadrons
from the geometry on the complex-$N_c$ plane.
Wide applications of the model to resonance physics are
expected in near future.

\section*{Acknowledgment}
This work is supported by 
Grant-in-Aid for Scientific Research on Innovative Areas
``Elucidation of
New Hadrons with a Variety of Flavors''
(Nos. 22105509 (K. N.), 
22105510 (H. N.), 24105706 (D. J.) and E01:21105006 (A. H.))
from the Ministry of Education,
Culture, Sports, Science, and Technology(MEXT) of Japan.
K. N. is supported by the Special Postdoctoral Research Program
of RIKEN.

\end{document}